# A Laboratory-Scale Experiment and a Numerical Simulation of Unusual Spiral Plumes in a High-Prandtl-number Fluid

A.N. Sharifulin, V.A. Bezprozvannikh[1], A.N. Poludnitsin[2]

Perm National Research Polytechnic University, 614990 Perm, Russia

[1] Fluid, Heat & Electronics, #406 9637 108 Ave. NW, Edmonton, Alberta, Canada T5H4G4

[2] Perm State National Research University, 614990 Perm, Russia

**Abstract** We experimentally and numerically investigated the generation of plumes from a local heat source (LHS) and studied the interaction of these plumes with cellular convective motion (CCM) in a rectangular cavity filled with silicon oil at a Prandtl number (Pr) of approximately two thousand. The LHS is generated using a 0.2-W green laser beam. A roll-type CCM is generated by vertically heating one side of the cavity. The CCM may lead to the formation of an unusual spiral convective plume that resembles a vertical Archimedes spiral. A similar plume is obtained in a direct numerical simulation.

We discuss the physical mechanism for the formation of a spiral plume and the application of the results to mantle convection problems. We also estimate the Reynolds (Re) and Rayleigh (Ra) numbers and apply self-similarity theory to convection in the Earth's mantle. Spiral plumes can be used to interpret mantle tomography results over the last decade.

**Introduction** This study was inspired by the important role played by the thermal convection of large-Prandtl (Pr)-number fluids in geophysics. The investigation of convective plumes that are generated by point sources can be used to model heat and mass transfer in the Earth's mantle. The mantle material is characterized by an extremely high Pr, i.e. Pr ≈ $10^{-23}$ (Golitsin (1979)). Within the currently accepted theory of mantle convection, the generation of mantle plumes by powerful heat sources at the border of the Earth's core is used to explain many important geological features and processes.

We experimentally simulate the generation of a convective plume from a hot spot and study its interaction with the cellular flow of a high-Pr fluid. Direct CFD simulations are used to model the experimental results.

**Experimental procedure** We use a silicone oil PMS-200 with a density 963 kg/m$^3$ and a Pr ≈ 2·10$^3$ (see Vargaftik (1972)). From our laboratory measurements, we estimate that the fluid has a volumetric thermal expansion coefficient of β≈10$^{-3}$ 1/°C and a kinematic viscosity of ν≈2·10$^{-4}$ m$^2$/c. A convective plume is generated by a local heat source (LHS) on the top of a rubber cylinder, which is placed at the center of the bottom of a rectangular cell (see Fig. 1). A green laser with a maximum power of 0.2 W is used to simulate a hot spot. The angle of incidence of the laser light ranges from 56° to 61°. The average oil temperature in the cavity is near 30°C. A thermocouple is placed in the hot spot and shows that the fluid temperature exceeds the temperature of the hot spot by δT=13°C at the maximum laser power. The laser beam enters through the transparent left vertical edge of the cavity. A 100 W electric heater replaced in 1 m generated infrared radiation and produces a slight warming effect, such that there is a temperature difference of 0.01°C≤ ΔT ≤0.03°C between the left and right edges of the cell in all of the experiments. The cell has an internal volume of 90 x 90 x 80 mm$^3$. The front and rear walls of the cell are made of optical glass through which the flow is observed. The other sides of the cell are composed of plastic. All of the walls have a thickness of 10 mm. The cell is filled with oil up to a height of 83 mm. These conditions create a cellular steady-state flow with a velocity of approximately 0.05 mm/s near the free surface of the fluid.

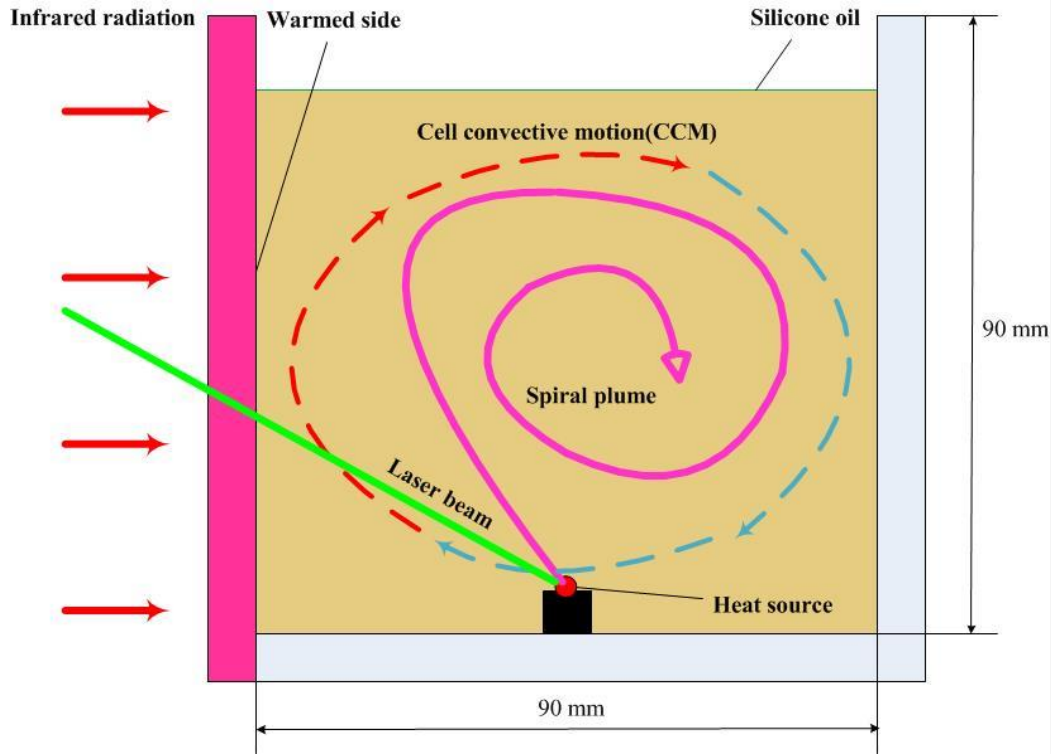

**Fig. 1** Schematic of experimental configuration (see description in text)

The flow visualization is performed using the Schlieren method with a IAB-451 device at a frame rate of 20 s. The intensity of the convective motion in the plume is estimated by calculating the Grashof number Gr=gβd³δT/ν² and the Rayleigh number Ra=Gr·Pr. The maximum laser power corresponds to $Gr_m=0.5$ and $Ra_m=10^3$, at which the plume is straight and vertical with a diameter in the area of the heat source of d≈10 mm). The plume reaches the free surface of the fluid in 700 s, as has been typically observed in previous studies (see Kumagai (2002) and Kaminski and Jaupart (2003)). The results presented here are obtained for a beam power of 0.12 W. The excess temperature at the hot spot is δT≈2ºC. After switching on the laser, a thin cylindrical plume appears (see Fig. 2) and grows in the direction of the cellular flow. The plume breaks down at approximately half of the cavity height and then drifts with the cell convective motion (CCM), forming a spiral. The height of the oil above the heat source is 7.2 cm. The first turn of the spiral occurs at a height of 4 cm, and the uppermost point of the spiral coil is located at a height of 6 cm. The upper part of the plume is almost horizontal. As the number of turns of the spiral increases, this horizontal section rises and approaches the free surface. The dynamics of the plume that is generated above a large rubber cylinder are visualized as a movie that is available as online supplementary material.

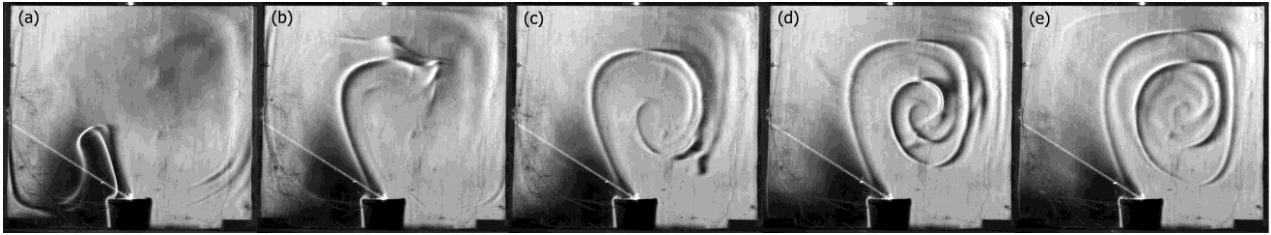

**Fig. 2** Shape of plume generated above a small rubber cylinder at five time points $t$ after switching on the laser: (a) $t = 980$ s; (b) $t = 3084$ s; (c) $t = 5688$ s; (d) $t = 9852$ s and (e) $t = 13505$ s; plume growth rate is approximately $u_p = 0.04$ mm/s; and cell convective motion velocity on the upper free surface is $u_{ccm} = 0.05$ mm/s

The Gr and Ra numbers for a vertically growing plume with a diameter of d≈4 mm are estimated at $Gr_p$=0.03 and $Ra_p$=0.6·$10^2$, indicating that the fluid in the plume is in a laminar and stable flow. Assuming that the temperature difference between the vertical walls is $\Delta T \approx 0.01$°C, we obtain the Gr and Ra for the CCM, i.e., $Gr_c$=2 and $Ra_c$=4·$10^3$. For large Pr, the temperature acts as a passive tracer. There is a negligible amount of heat transfer between the CCM and the plume, resulting in a long-lived plume. The momentum transfer between the CCM and the plume is not negligible at the initially straight section of the plume. The fluid momentum in this area of the plume is almost completely in the vertical direction, and the CCM velocity is almost completely is the horizontal direction. Therefore, the horizontal component of the momentum diffuses into the plume, which causes the plume to bend. However, the vertical component of the velocity is transferred from the plume to the CCM, resulting in an asymmetry between the top and bottom of the CCM.

**Numerical simulation** The ANSYS Fluent software suite is used to numerically simulate the unusual liquid plume phenomenon that is observed in the aforementioned laboratory-scale experiment. We consider that the dominant heat transfer mechanism is natural convection for which the temperature differences over the domain are small. Thus, we can use the Boussinesq approach, which exhibits faster convergence for many buoyancy driven flows than conventional models in which the fluid density is considered temperature-dependent. The Ra is also sufficiently small in this model that the flow can be considered to be laminar. The three-dimensional continuity, momentum, and energy equations are solved to simulate the incompressible convective flow of the transformer oil that is used in our experiment. Steady-state calculations were performed using the Boussinesq model. Note that steady-state computations may be useful for routine technical applications (such as for evaluating the heat transfer in transformers and electronic components) and are applicable to a variety of currently controversial physics research problems, such as the behavior of the Earth's mantle (see below). The computational domain corresponds to half of the experimental cavity because of the symmetry of the problem. The computational grid consists of 711696 cells, 2160836 faces, and 737730 nodes. The following thermal boundary conditions are applied: the temperatures of the bottom, front, back, and right walls, and all of the heated cylinder walls at the bottom of the chamber (except for the top surface) are 300 K. The temperature of the left wall (through which the laser beam enters) is 300.5 K. Symmetry conditions are applied at the front of the domain, and outflow conditions are applied at the free surface of the fluid. The thermal conditions for two of the laboratory experiments are used in the computations. The laser heats the top surface of the internal cylinder to temperatures of 302 K and 312 K. The PRESTO scheme was used to discretize the pressure, and the second order upwind scheme was used to discretize the momentum and energy equations. A SIMPLE scheme was used to discretize the pressure-velocity coupling equation.

The numerical simulations produce plume-like flow structures that are similar to those observed in the physical experiments. Figures 3 and 4 illustrate the pathlines in the symmetry zones of the domain that are emanate from the top surface of the heated inner cylinder for heat source temperatures exceeding 12 K and 2 K, respectively.

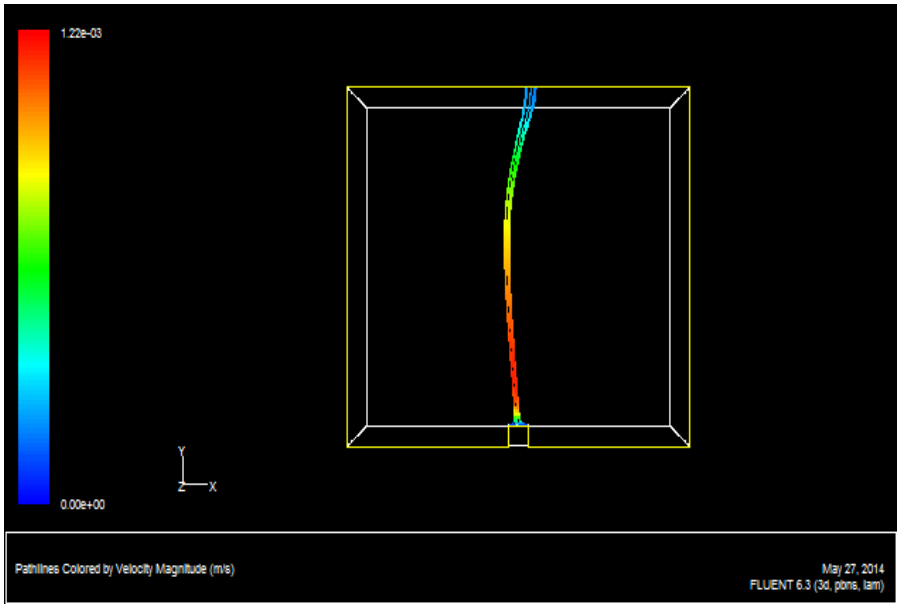

**Fig. 3** Pathlines for temperatures exceeding 12 K: different colors indicate varying magnitudes of velocity

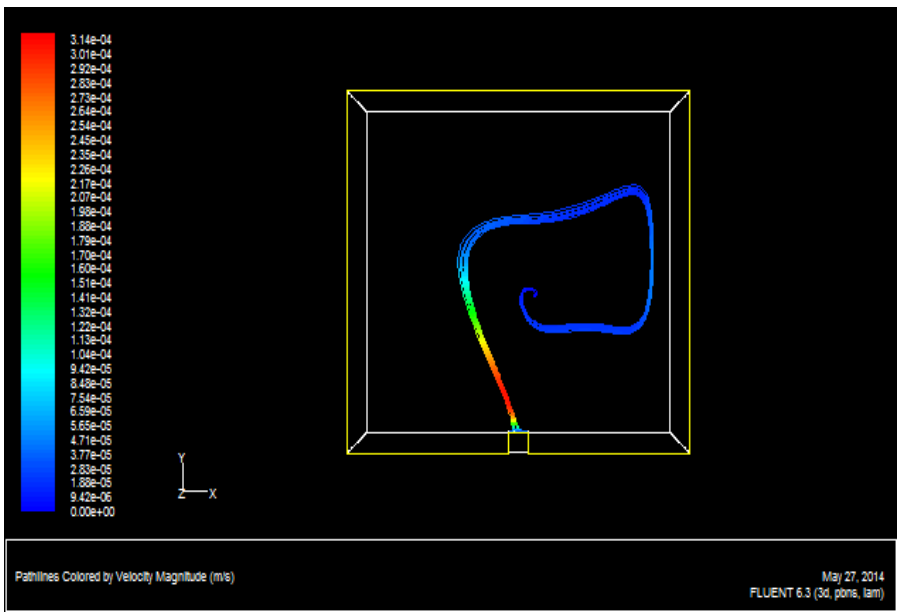

**Fig. 4** Pathlines for temperatures exceeding 2 K: different colors indicate varying magnitudes of velocity

Thus, the numerical simulations reproduce the unusual spiral plume that is observed in the physical experiments. This result validates the applicability of the CFD approach to simulate natural convection effects in all of their complexity. For the higher surface temperatures, the results diverge because of the use of the steady-state assumption in the computations.

**Application to convection in the Earth's mantle** Golitsin (1979) found self-similar convection solutions at low Re and high Pr, i.e., Re << 1, Pr >> 1. This result provides a physical basis for relating our laboratory-scale simulation, including the modeling of the flow patterns, to thermal convection in the Earth's mantle. The Re=u·d/ν for the vertically growing plume, with u=$u_p$, is estimated at $Re_p$ ~ $10^{-3}$. The Re for the CCM is also small, $Re_{ccm}$=0.02, and the Pr is high. Thus, the conditions for the laboratory experiment are close to fulfilling Golitsin's conditions for self-similarity.

There are over 100 regions of long-lived and extensive volcanism (hot spots) over the surface of the Earth. Most of these hot spots occur at plate boundaries and are directly associated with plate tectonics. Numerous hot spots of various sizes are found around the world. The locations of these hot spots are fairly stable with respect to each another, but their relative motion averages several millimeters per year, which is less than 10% of the speed of a tectonic plate: $u_p \approx 10\ cm/year$. The remarkable mutual immobility of these hotspots led researchers in the 1960s to directly associate these hot spots with the Earth's hot core and to formulate the plume hypothesis (see Humphreys and Schmandt (2011) and Ribe et al. (2007) for further details and references). Advances in mantle tomography over the last decade have made it possible to observe sections of locally heated regions (LHRs) under hot spots. Montelli et al. (2006) reported on horizontal sections below 60 hotspots, and Zhao (2007) reported on vertical sections near 30 hotspots. The studies show a chaotic distribution of LHRs over these sections. The quantity and distribution of the LHRs in the horizontal sections varies widely with the depth of the section. Thus, if plumes indeed exist in the mantle, they must have a very different shape from the vertical heat columns that have been described in previous models of mantle convection. However, the chaotic distribution of LHRs could be explained by mantle plumes with spiral shapes that are similar to those observed in our experiments. For example, the horizontal sections of the spiral that are presented in this paper exhibit one, two, three and four LHRs. Gradual increases in the depth of the predetermined horizontal section lead to a change in the shape of a LHR and consequently, a new LHR or to the sudden disappearance of the existing LHR. The horizontal sections for the first LHR and the corresponding section of the tangent of the upper loop of the spiral can be highly elongated, as has been observed in the hot spot area of Yellowstone (see Humphreys and Schmandt (2011)). The spiral shape of the plume can also explain the sharp decrease in the number of LHRs when approaching the Earth's core.

**Conclusions** We have presented an experimental and numerical study on plumes that are generated by a LHS in a large-Pr fluid with cellular convection. If the magnitude of the velocity of the initial plume is close to that from convection motion, the plume forms an unusual spiral shape.

**Acknowledgments** The authors would like to thank Professor Victor I. Stepanov from the Institute of Continuous Media Mechanics for providing essential materials and sharing his insights into this problem.